\documentclass[preprint,11pt]{elsarticle}

\usepackage{setspace}
\usepackage{graphicx}
\usepackage{amssymb}
\usepackage{xcolor}
\usepackage{ulem} 
\usepackage{algorithm}
\usepackage[noend]{algpseudocode}

\usepackage{adjustbox}

\usepackage{lipsum}

\usepackage[compact]{titlesec}         
\titlespacing{\section}{0pt}{0pt}{0pt} 
\AtBeginDocument{
  \setlength\abovedisplayskip{0pt}
  \setlength\belowdisplayskip{0pt}}
  
\makeatletter
\def\BState{\State\hskip-\ALG@thistlm}
\makeatother

\usepackage{pdflscape}

\usepackage{graphicx}

\usepackage{tikz}

\usepackage{mathabx}
\def \cind {\perp\!\!\perp}

\usepackage{hyperref}

\journal{Journal Name}

\begin{document}

\begin{frontmatter}


\title{More  investment in  Research and Development  for better Education in the future?}


\author{Rim Lahmandi-Ayed and Dhafer Malouche}

\address{ESSAI and MASE-ESSAI, University of Carthage }

\begin{abstract}
The question in this paper is whether R\&D efforts affect education performance in small classes. Merging two datasets collected from  the PISA  studies  and the World Development Indicators and using Learning Bayesian Networks, we prove  the existence of a statistical causal relationship between investment in R\&D of a country and its education performance (PISA scores). We also  prove that the effect of R\&D on Education is  long term  as a country has to invest at least 10 years  before beginning to improve the level of young pupils.

\end{abstract}


\end{frontmatter}


\section{Introduction}
\label{S:1}
Education is considered among the basic human rights all over the world. This is  not  surprising after the abundant literature recognizing the numerous benefits of education. Besides being  desirable in itself, education  has positive effects at the individual level in terms of wealth, health, happiness within the couple and the family; favors social cohesion   reducing crimes and raising voluntary activities; and additionally has positive   economic benefits,  increasing productivity, economic growth and competitiveness. 

On the one hand, it is  necessary, given the importance of education in human life, to study the determinants of education performance.  On the other hand, numerous studies exist to measure the short-term effects of Research and Development. But  ``we know surprisingly little about their long-term effects. This is a pity, because the conventional justifications for state intervention in research depend upon phenomena  that are inherently long term in nature.'' (\citeauthor{Arnold2012}, \citeyear{Arnold2012}) To the best of our knowledge, there is no study linking R\&D and education although both are components of the same educative system.

In this paper, we argue that R\&D expenditure  affects positively education in the long run. More precisely, we  prove the existence of a statistical causal relationship between investment in R\&D and education performance (PISA scores). We also measure the lag between the R\&D funding and its effect on education performance and prove that this effect is  long term  as a country has to invest at least 10 years  before beginning to improve the level of young pupils.


 The scarcity of studies of long-term effects of R\&D, among which the effect on education performance, is due to the fact that the stakeholders are only concerned by effects expected in a horizon relevant to their elective life or budgeting cycle\footnote{This is the so called ``dynamic inconsistency" existing between the long time constants of R\&D and shorter cycles in political, administrative and budgeting terms referred to by \citeauthor{Arnold2012} (\citeyear{Arnold2012}).}. But because long term effects of R\&D are hardly studied, the decisions taken concerning R\&D may be sub-optimal ones, as they do not take those effects into account. 
 
 Measuring the effect of R\&D expenditure on education performance in small classes and how long we must wait for these effects, is very important in general. It is  more important in particular in developing countries where  research is not given much importance, believing that they can do without research by benefiting from the research of others through ready-for-use items and technologies, and choosing to focus on education in primary and secondary schools. But investment in  R\&D affecting education of children, decision-makers must understand that if R\&D is neglected, not only researchers and universities will suffer but  the performance of primary and secondary schools  is also at stake, which may threaten the whole educative system.  


 Our proof in this paper is based on a statistical modeling performed from merging  two datasets collected from two different sources: the PISA  studies made by the OECD and the World Development Indicators provided by the World Bank.
 
 ``The Programme for International Student Assessment (PISA) is a triennial international survey which aims to evaluate education systems worldwide by testing the skills and knowledge of 15-year-old students'' in five disciplines\footnote{http://www.oecd.org/pisa/}.
 We consider the 2015 study, the most recent one, to benefit from the maximal hindsight, focusing on the national scores  obtained in mathematics, reading and science which  are our \textit{output} variables. 
 
 The second  dataset used in this paper is provided by the World Bank  in the so-called World Development Indicators available on the web and in a package from the  \texttt{R} statistical software\footnote{\url{r-project.org}}. This database contains several indicators by country about R\&D  from which we compute a new R\&D indicator defined as the amount of expenditure per researcher.  
 


We use Learning Bayesian Networks to estimate the relationships in the random vector constituted by the annual R\&D expenditure  from 1997 to 2014 and the PISA scores in 2015 (\citeauthor{Pearl2000}, \citeyear{Pearl2000}; \citeauthor{SpirteGylmour1991}, \citeyear{SpirteGylmour1991}; \citeauthor{scutari2015}, \citeyear{scutari2015}). Before making the estimation, the Bayesian Networks are also used to impute the missing values, thus providing  an imputation method better than the existing ones. 

Noting that the three considered scores (reading, mathematics and science) are strongly correlated we restrict our study to reading only to avoid redundancy. We then measure the strength of the link between each annual R\&D expenditure and the reading score. We prove that the investment  the most likely linked to  the ``Reading'' variable is the one made in 2005. This means that the efforts made in R\&D begin to show results only 10 years later. 

As an interpretation of the obtained results, we think that R\&D affects education in small classes through the training of trainers and all types of educators who will be in charge of  children including  future parents. The money spent in R\&D raises the quality of university teaching, thus the quality of the graduates among whom the future adults in charge of children. This explains in the same time why the delay is long. Indeed the effect of R\&D on education is not a direct one. It occurs through a chain of effects, each needing some delay. A dollar spent in R\&D has first  to impact the performance of researchers (publications, patents...), which in turn impacts the quality of university teaching at the individual level and through a richest  more dynamic university  environment. Better university teachers  give better graduates, so better future parents and educators, which finally impacts the education performance.

The delay of 10 years is  too long  for countries under financial stress and is beyond the horizon relevant to the elective life or budget cycles in democracies. Cutting the research budgets is almost a reflex of decision-makers at each crisis, this must not be surprising. But this reflex is a bad one, an irresponsible decision as it undermines the education of the future generations. What happens in this respect is very similar to environment issues where decision-makers may take decisions for which they do not have to bear the consequences.

As a by-product of the estimation, we  measure the contribution of R\&D in the explanation of the performance of Education in each country. This contribution ranges from 45\% to 64\%, which is very high. This means that R\&D contributes highly in the performance of Education.

This conclusion is however mitigated by further results. Indeed, we also evaluate the efficiency of R\&D in the performance of Education obtained from the comparison of the estimated performance of education and the observed value. Related to the contribution of R\&D in the explanation of Education which varies highly across countries, R\&D contributes more or less in the performance of Education because countries invest more or less in R\&D and/or because R\&D is more or less efficient. This is to say that investing more in R\&D may be insufficient. Budgets have to be more efficiently spent and R\&D more efficiently organized in order to secure the transmission of the benefits to education performance.

{\it The literature review.}  We have first to make clear that we deal with research in general and  not particularly  with  research in educational sciences  as is done by \citeauthor{Cooper53} (\citeyear{Cooper53}), \citeauthor{levin2004} (\citeyear{levin2004}) and certainly numerous others. 

We can relate our paper to an abundant literature, empirical and theoretical, on the socio-economic impact of knowledge in a broad sense: schooling and basic education (\citeauthor{haveman84}, \citeyear{haveman84}), Higher education (\citeauthor{ocaroll2006}, \citeyear{ocaroll2006}) and research and development  (\citeauthor{beck2017}, \citeyear{beck2017}).

There are numerous studies on economic effects of R\&D: productivity (\citeauthor{griliches79}, 1998 and \citeauthor{Lichtenberg1991}, \citeyear{Lichtenberg1991}), technological change (\citeauthor{Schumpeter1942}, \citeyear{Schumpeter1942}) which is the source for sustainable economic growth (\citeauthor{Solow1956}, \citeyear{Solow1956}). Because of spillover effects of the knowledge creation between firms (\citeauthor{beck2017}, \citeyear{beck2017}; \citeauthor{hmm10}, \citeyear{hmm10}; \citeauthor{bloom13}, \citeyear{bloom13}; \citeauthor{Cardamone12}, \citeyear{Cardamone12}; \citeauthor{venturini2015}, \citeyear{venturini2015}; \citeauthor{Acharya2015}, \citeyear{Acharya2015}; \citeauthor{bloch2013}, \citeyear{bloch2013}), the market may fail in providing the optimal level of R\&D investment, which justifies   public funding of R\&D (\citeauthor{MartinScott2000}, \citeyear{MartinScott2000};  \citeauthor{Romer1990}, \citeyear{Romer1990}; \citeauthor{jw98}, \citeyear{jw98}).  

 Numerous papers are interested in the effects of public  and university research.
 \citeauthor{jth93} (\citeyear{jth93}), \citeauthor{jt1996} (\citeyear{jt1996}),
 \citeauthor{mohnen1996} (\citeyear{mohnen1996}),  \citeauthor{Blomstrom98} (\citeyear{Blomstrom98}), \citeauthor{Cincera01} (\citeyear{Cincera01}) study the role of university R\&D in increasing productivity and improving the competitiveness of countries. 
Papers identify the importance of academic research in driving economic growth (\citeauthor{hc97}, \citeyear{hc97}) and regional economic development (\citeauthor{smilor1993}, \citeyear{smilor1993}). Science parks or innovation centers most commonly located on university campus and facilitating many spillovers and benefits from the proximity with researchers and innovation, help new companies flourish, which is a channel to create jobs and foster economic growth   (\citeauthor{ocaroll2006}, \citeyear{ocaroll2006}).

A stream of literature deals with the socio-economic effects of basic education (schooling or basic education completion) and/or higher education (access or completion). The theoretical papers of \citeauthor{barro91} (\citeyear{barro91})  and \citeauthor{lucas88} (\citeyear{lucas88})  and the empirical study of \citeauthor{Dension85} (\citeyear{Dension85})   demonstrate the positive role of education (or more generally human capital formation) in economic growth. \citeauthor{moretti2004} (\citeyear{moretti2004}) establishes that an increase in the rate of graduates over the population of workers increases the wages for all workers. \citeauthor{shultzs61} (\citeyear{shultzs61}),  \citeauthor{hansen63} (\citeyear{hansen63}),  \citeauthor{Becker1964} (\citeyear{Becker1964}), \citeauthor{mincer62} (\citeyear{mincer62})  among others have earlier studied the relation between productivity and schooling.

A body of literature deals with non-marketed effects of education. First education may be consumed for its intrinsic value (\citeauthor{Lazear77}, \citeyear{Lazear77}).
 Several studies exist on the relation between higher education and voluntary activities: \citeauthor{Freeman97} (\citeyear{Freeman97}), \citeauthor{vaill1994} (\citeyear{vaill1994}), \citeauthor{gibson01} (\citeyear{gibson01}), \citeauthor{Dee03} (\citeyear{Dee03}). \citeauthor{Dye80} (\citeyear{Dye80}) and \citeauthor{mueller1978} (\citeyear{mueller1978}) have shown earlier that education increases both money and time donations. Several papers examine the relation between education and health issues: \citeauthor{grossman75} (\citeyear{grossman75}), \citeauthor{fuchs82} (\citeyear{fuchs82}), \citeauthor{leigh81} (\citeyear{leigh81}),  \citeauthor{Farrell82} (\citeyear{Farrell82}), \citeauthor{k91} (\citeyear{k91})  and \citeauthor{llm2005} (\citeyear{llm2005}).  \citeauthor{Ehrlich75} (\citeyear{Ehrlich75}) has  shown that education reduces criminal activity and \citeauthor{locmor2001} (\citeyear{locmor2001}) investigate the relationship between High School completion and crime. Numerous papers establish a positive correlation between education and attainment of desired family size: \citeauthor{michael73} (\citeyear{michael73}), \citeauthor{ryder1965} (\citeyear{ryder1965}), \citeauthor{mw76} (\citeyear{mw76}); and a positive correlation between education and sorting in the marriage market (\citeauthor{Becker77}, \citeyear{Becker77} and \citeauthor{j69}, \citeyear{j69}).  Importantly for our purpose, numerous papers have proved that mother's and father's education positively influences child quality in several respects (health, cognitive development, education, occupation status, future earnings): \citeauthor{l75} (\citeyear{l74}, \citeyear{l75}), \citeauthor{Edwards79} (\citeyear{Edwards79}), \citeauthor{bg70} (\citeyear{bg70}), \citeauthor{hs74} (\citeyear{hs74}, \citeyear{hs80}), \citeauthor{wb82} (\citeyear{wb82}), among numerous others.

To the best of our knowledge, no paper deals with the effect of R\&D expenditure on education performance. In a broader sense, there is no paper dealing with the determinants of education performance. However there are numerous papers dealing with the determinants of research and innovation.   Berman
(1990) establishes that university-funded research stimulates the industrial R\&D. Numerous papers deal with the effects of  direct and/or indirect public  funding for private R\&D on the firms' efforts in R\&D: \citeauthor{dp16} (\citeyear{dp16}),   \citeauthor{hh15} (\citeyear{hh15}), \citeauthor{d04} (\citeyear{d04}) among numerous others.  Science parks and innovation centers foster industry oriented innovations through the transfer of academic research. Finally some old papers  backward trace innovations of interest and identify the scientific events that fed into them: \citeauthor{i67} (\citeyear{i67})  on the weapons systems, \citeauthor{l68} (\citeyear{l68}, \citeyear{l69}) on five well-known innovations (such as the videotape recorder and oral contraceptives).

Finally, our paper may be related to \citeauthor{pillis2001} (\citeyear{pillis2001}) who develop a mathematical model to explore the long-term effects of university funding on: the population of active professors, the number of students and the creation of jobs in the industry.

The remainder of the paper is organized as follows. Section 2 describes the data. Section 3 presents the methods used in the imputation and in the estimation of the model. Section 4 provides the results. Section 5 concludes.
   

\vskip 5 truemm
\section{Data design\label{DataDes}}
\vskip 5 truemm
In this paper we  design a dataset from two very well known sources. The first source called PISA data is collected from the Program for International Student Assessment (PISA).
This is a triennial international survey which aims to evaluate education systems worldwide by testing the skills and knowledge of 15-year-old students\footnote{\url{http://www.oecd.org/pisa/}}. Every three year period, beginning from 2000, around one half million students from over 72 countries are tested in science, mathematics, reading, collaborative problem solving and financial literacy. In this paper we focus on the performance in mathematics, reading and science obtained in the 2015 study. We use the most recent PISA study in  order to benefit from the maximal hindsight.

The  raw data are individual marks obtained by  the students in each country. This data is  available in either a SAS or SPSS format (see \url{http://www.oecd.org/pisa/data/2015database/}). They are large data files, around 400 MB each. Once downloaded, we  use the R package \texttt{intsvy} (see \citeauthor{JSSv081i07}, \citeyear{JSSv081i07}) to compute the national scores in mathematics, science and reading   obtained as the average of the individual marks within each country. 
 These scores are our \textit{output} variables. Denote by $\mathcal{Y}$ the obtained series of observations of the PISA national scores.

The second data is obtained from  a database  known as the World Development Indicators (WDI) collected by the World Bank. It consists in  a grouping of 800 indicators covering more than 150 economies. They are compiled from officially recognized national sources. This data presents the most recent and accurate global development data available, and includes national, regional and global estimates\footnote{\url{https://datacatalog.worldbank.org/dataset/world-development-indicators}}. This data is  downloaded and managed with the \texttt{R} package  \texttt{WDI}\footnote{\url{https://github.com/vincentarelbundock/WDI}}. 
The R\&D indicators contained in the WDI database are available only from 1997 to 2014 with some reasonable missing values. This is why we limit  our study to this  period.

To measure the investment of countries in R\&D, we choose to consider the annual expenditure per researcher. Indeed the percentage of GDP granted to R\&D cannot be the right variable as it varies very little among countries. Neither is the gross amount granted to R\&D as it varies too much and must be put in perspective   relative to the number of researchers. The  annual expenditure per researcher appears to be a reasonable choice with a reasonable variance. We have to calculate it as it is not available directly in the database. 
To do so, among the several indicators related to R\&D activities of the countries, we use the two following indicators:

\begin{itemize}
\item \textbf{Expend}: Research and Development expenditure (\% of GDP). It is defined as the amount in dollars of the Expenditures for Research and Development. It includes the current and capital expenditures (both
public and private) on creative work undertaken systematically to increase knowledge, including knowledge of humanity, culture, and society, and the use of knowledge for new applications. R\&D covers basic research, applied research, and experimental development.

\item \textbf{NumbRD}: Number of researchers. It is  defined as the number of Researchers per million people. The researchers are defined to be the professionals engaged in the conception or creation of new knowledge, products, processes, methods or systems and in the management of the projects concerned, including postgraduate PhD students.

\end{itemize}
We  also extract from the WDI data  two other indicators:  the GDP in dollars (\textbf{GDP}) and the total size of the population (\textbf{Pop}). This enables us to compute the  two following  indicators:

\begin{itemize}
\item Total amount of Expenditure in dollars
\begin{equation}
\label{defTotExp}
\mathbf{TotExp}= \mathbf{Expend} \times \mathbf{GDP} \times 10^{-2}.
\end{equation}
\item Total Number of Researchers
\begin{equation}
\label{defTotRD}
\mathbf{TotRD}= \mathbf{NumbRD} \times \mathbf{Pop} \times 10^{-6}.
\end{equation}

\end{itemize} 

We thus obtain the Expenditure per Researcher 
\begin{equation}
\label{defTotRD2}
\mathbf{ExpOneRD}= \displaystyle\frac{\mathbf{TotExp}}{\mathbf{TotRD}},
\end{equation}
and  a time series for each country defined by the annual expenditure that a given country  spends for the activity of one researcher from  year 1997 to year 2014:
$$ \mathbf{ExpOneRD}(t),\; \;t=1997,\ldots,2014.$$

\begin{figure}
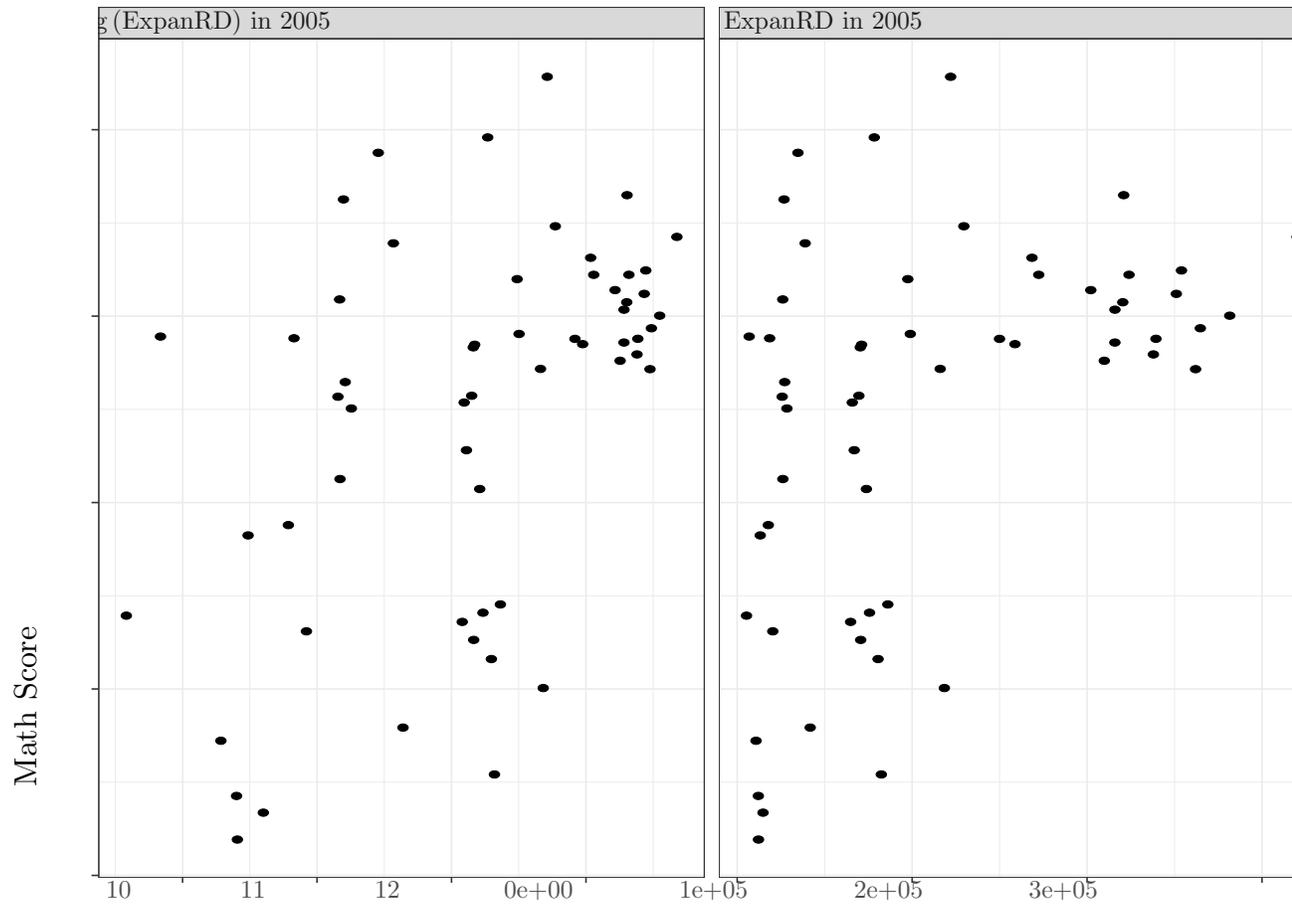

\centering


\caption{Two scatter plots: (a) $\log\mathbf{ExpOneRD}$ in 2005 $\times$ Math Score, (b) $\mathbf{ExpOneRD}$ in 2005 $\times$ Math Score,  \label{whylog}}
\end{figure}

As will be better justified later, it is more appropriate to use a logarithm transformation of the $\mathbf{ExpOneRD}(t)$ variables instead of using them in a raw format. Indeed when we look at the relationship between the $\mathbf{ExpOneRD}$ variables and the PISA scores we notice that above a certain amount of investment, the PISA scores vary very little. There is indeed a kind of bearing in the scatter plot (see Figure \ref{whylog}) suggesting rather a logarithmic relationship between the $\mathbf{ExpOneRD}$  variable and the PISA scores. 

Hence, our final \textit{input} data, denoted by $\mathcal{X}$ and called  the RD investment data, is composed of the variables  equal  to the logarithm transformation of  $\mathbf{ExpOneRD}(t)$:
$\mathcal{X} = \{ \mathbf{\log (ExpOneRD}(t)), \;t=1997,\ldots,2014\}$.

The last Data design step consists in merging $\mathcal{X}$ and $\mathcal{Y}$, the two recently constructed datasets, in order to obtain one data containing the  PISA national scores of 2015  (either Math or  Reading or  Science) and the  RD investment data  $\mathcal{X}$.  This restricts our analysis to the 57 countries or economies belonging in the same time to both datasets. Our  data can be denoted by
\begin{equation}
\mathcal{D}=\left[\mathcal{Y},\,\mathcal{X}\right].
\label{rawData}
\end{equation}

But because of the missing values in the raw data $\mathcal{X}$, we need  to go through    a process of data cleaning and imputation. 
This procedure is  somewhat innovative and also based on the estimation of Bayesian Networks  as we will explain in the next section.


\vskip 5 truemm
\section{Methods }
\vskip 5 truemm
We have then constructed  a data $\mathcal{D}$ that can be considered as an $n-$sample of observations of the random vector 
\begin{equation}
\label{randVect}
[Y,\mathbf{X}]=\left[Y,(\,X_{1997},\ldots,X_{2014})\right],
\end{equation}
where $Y$ is a random variable representing one of the PISA scores (Math, Reading or Science)  and $X_{1997},\ldots,X_{2014}$ are the random variables representing the $\log (\mathbf{ExpOneRD})$ variables. Our objective in this paper is to estimate the probability distribution of the random vector $[Y,\mathbf{X}]$.

To do so, we  estimate the model as a Bayesian Network using Gaussian Bayesian Networks called Learning Bayesian Networks. We first describe Bayesian Networks, then the estimation process and finally the imputation method also using Bayesian Networks.

Bayesian Networks (BN)  are Directed Acyclic Graphs (DAG) used to read the relationships between the variables in the random vector $[Y,\mathbf{X}]$. Mathematically speaking a BN is  a couple $G=\left( V, E \right)$ where $V$
is the set of nodes.  Each node $v\in V$ represents one variable from the random variables associated to the input or output data $\{Y\} \cup \{X_t, t= 1997,\ldots,2014\}$, and  $E$ is the set of directed edges. 

$E$ is  a subset of $V \times V$ such that, if $(v,v')\in E$ then $(v',v)\not\in E$.  An element $(v,v')$ in $E$ is then denoted by  $(v\rightarrow v')$. Thus the elements of $E$ are arrows with  directed edges. In this paper we will denote for every $v\in V$ by $\theta(v)$ the variable in $[Y,\mathbf{X}]$ represented by the node $v$ in the DAG $G$.  $\theta(v)$ can be either $Y$ (the PISA score) or one of the variables $X_t$ (the investment variables in R\&D).  If $S \subseteq V$, the sub-random vector indexed by $S$ in  $[Y,\mathbf{X}]$  will be then denoted by $\Theta(S)=(\theta (s),\,s\in S)$.

If $(v\rightarrow v')\in E$ then $v$ is called the \textit{parent} of $v'$ and $v'$  the \textit{child} of $v$. A \textit{path} from an  edge $v$   to another edge $v'$ is a sequence of edges $v_0,\ldots,v_n$ such that $v_0=v$, $v_n=v'$ and for every $i=0,\ldots,n-1,$ 
$(v_{i}\rightarrow v_{i+1})\in E.$

The DAG $G$, when associated to a random vector $[Y,\mathbf{X}]$, is then used to read the conditional independence between the variables in the random vector $[Y,\mathbf{X}]$. In fact $G$ helps to give a parsimonious factorization of the probability density of the random vector $[Y,\mathbf{X}]$. Hence if $f$ is the density of $[Y,\mathbf{X}]$, this density may  factorize when $G$ is known, i.e. may take the   following form:
\begin{equation}
\label{facMark}
f(\Theta)=\displaystyle\prod_{v\in V} g(\theta(v)\mid \Theta(pa(v))),
\end{equation}
where $\Theta=(\theta(v),\,v\in V)\in \mathbb{R}^{|V|}$ is the random vector $[Y,\mathbf{X}]$ and $pa(v)$ denotes the set of parents of the node $v$ in $G$. 

Equality  (\ref{facMark}) is called the factorization Markov property of the probability of $[Y,\mathbf{X}]$ with respect to $G$. When  $(\ref{facMark})$ is satisfied we can also conclude that for any couple of nodes $v$ and $v'$ the following pairwise Markov property is satisfied:

\begin{equation}
\label{pairMark}
\mbox{If  neither}\; (v\rightarrow v') \mbox{ nor }(v\rightarrow v')\in E\;\mbox{ then }\; \theta(v)\cind \theta(v')\mid \Theta(pa(v)),
\end{equation}
where $\theta(v)\cind \theta(v')\mid \Theta(pa(v))$ means that the variables $\theta(v)$ and $\theta(v')$ are independent given the variables of the random vector $\Theta(pa(v))$.

Hence from a statistical point of view when the data is  generated from variables satisfying the Factorization Markov Property as in (\ref{facMark}) the variables represented by the nodes not belonging to the set of parents of $v$ have no impact on the variable $\theta(v)$. This   was our motivation for the use of BN to study the effect of the investment in  R\&D on  Education. By estimating the BN from the data $\mathcal{D}$ we can not only know which  year's investment has a direct impact on the PISA scores but we also know the whole mechanism through which investments in R\&D affect the PISA scores. These relationships can  also be easily visualized using the representation of the DAG.

\subsection{Learning Bayesian Network}

To achieve our objective we need  to estimate from our collected data $\mathcal{D}$ the Bayesian Network  $G$ as defined above. Since this data is exclusively composed of the observation of quantitative variables, we   use Gaussian Bayesian Network to model the relationship between the PISA score $Y$ and the investment variables $X_t,\,t=1997$ to $t=2014.$  This process of estimation of $G$ is usually called the Learning structure of a Bayesian Network. Several algorithms exist in the literature for this  purpose. We usually  estimate the BN that corresponds to the  minimum of a score  measuring how  good the BN    fits  the data. The learning procedure is then a search of a minimum in a discrete space which is the set of possible DAGs. This problem is known to be an NP-hard problem and we have to use some searching algorithm adapted to the BN. Next we  cite some of them. 

First  notice that we are not  searching in the whole space of possible DAGs with nodes in $V$. In fact our DAGs should not contain edges from $X_{t'}$ to $X_{t}$ when $t'>t$,  or edges from $Y$ to any of the $X_t$ when $t$ and $t'$ belong to $\{1997,\ldots,2014\}$, as a variable at some year (investment in R\&D or the PISA scores in 2015) is not expected to influence the past. 
Hence from the investment variable in  year 1997 we can have 18 possible arrows or edges, i.e. 17 years from 1998 to 2014 and the Education variable, the variable 1998 will have 17 edges, and so on. So the cardinality of the possible BN that fits  our problem is:
$$ 18! = 18 \times 17 \times \ldots  \times 1 =6.402374 \times 10^{15},$$
which is a huge number.

This is why  sampling algorithms are necessary  in this kind of problems. In the literature, there are mainly two families of \textit{learning} Bayesian networks: Constraint-based Algorithms and Score-based Algorithms.

In the Constraint-based Algorithms, we may cite for example the  PC-algorithm  (\citeauthor{SpirteGylmour1991}, \citeyear{SpirteGylmour1991} and  \citeauthor{KalishBuhlmann2007},\citeyear{KalishBuhlmann2007}). It is implemented in the \texttt{pcalg} package in \texttt{R}. This algorithm has two steps. The first one consists in an estimation of the skeleton of the BN $G$ which is an undirected version of $G$. This estimation can be made using a successively pairwise conditional independence hypothesis testing. 

It is decided to remove the undirected edge between a pair of nodes $v$ and $v'$, when we can find a subset $S\subseteq V\setminus \{v,v'\}$ such that the null hypothesis 
$H_0\,:\, \theta(v)\cind \theta(v')\mid \Theta(S)$ is satisfied. 

This procedure can contain a huge number of hypothesis testing steps, the conditional subsets $S$ have either at maximum a fixed cardinality or they are contained in a subset of neighbors of $v$ and $v'$. The first step finishes  with an estimation of undirected graph called an estimation of the skeleton of $G$ and a set of conditional independence statements.  The second step consists in identifying the direction of the arrows. The set of conditional independence statements obtained in the first step is used to identify the direction of the arrows by using a basic notion in DAGs called V-structures (see \citeauthor{Pearl2000}, \citeyear{Pearl2000} and  \citeauthor{KalishBuhlmann2007}, \citeyear{KalishBuhlmann2007}).  The final result  is a partially directed graph with undirected  and directed arrows. We can then guess the direction of the undirected arrows using some kind of priors or information on the edges.

The second family of \textit{learning} algorithms is called \textit{Score-based Algorithms}. They are based on heuristic optimization techniques in order to search a maximum of a given score that measures the quality of fit. One of them is the   \textit{greedy search algorithms} such as Hill-Climbing with random restarts  (\citeauthor{Bouckaert95}, \citeyear{Bouckaert95}). They are based on a searching procedure starting from an initial BN, usually the  simplest one, and by adding or removing an edge until the score can no longer be improved\footnote{In \citeauthor{scutari2015} (\citeyear{scutari2015}) you can find on page 106 a detailed description of the \textit{hill-Climbing} algorithm (HC).}.

Finally there exist hybrid algorithms  which are a composition between constraint-based and score-based algorithms. We obtain algorithms like the Sparse Candidate algorithm (SC) (\citeauthor{fpn1999}, \citeyear{fpn1999}) and the Max-Min Hill-Climbing algorithm (MMHC) (\citeauthor{Tsamardinos06}, \citeyear{Tsamardinos06}).

In this paper we choose to use  the Hill-Climbing (HC) and MMHC algorithms since they are already implemented in the \texttt{R} package \texttt{bnlearn}\footnote{\url{http://www.bnlearn.com}} (\citeauthor{scutari2015}, \citeyear{scutari2015}).

The Hill-Climbing (HC) algorithm will provide at the end an estimation of the BN without any measure of the accuracy of this estimation. To obtain a measure of this accuracy we  use a Bootstrap procedure (\citeauthor{Efron1993}, \citeyear{Efron1993}). This is a very known technique in statistics  that allows assigning measures of accuracy and provides an  estimation of the sampling distribution of any statistics. Since we are here interested by the estimation of the presence or not of  directed arrows between pairs of nodes of $V$, by applying Bootstrap procedure we will be able to give a probability of the presence of an edge in the searched DAGs. The use of Bootstrap procedure in the estimation of BN is already implemented in \texttt{R} using the command \texttt{boot.strength} of the package \texttt{bnlearn}. We  use 500 of bootstrap replicates and use for every sample the Hill-Climbing algorithm. We obtain the strength of the presence of each edge among which we are especially interested in  the ones linked to the Education variable $Y$. First we have to see if this probability is high and then we consider  the highest probability as a threshold in order to compute an average  model   containing only the significant edges.  

On the one hand  these procedures can be  implemented only on complete data, i.e.  data with no missing values. On the other hand, as we explained earlier,  the investment data $\mathcal{X}$ has missing values. Hence we have first to go through a missing value imputation procedure to complete the missing observations before estimating our model. 
To do so, we implement a new procedure based again on the  learning structure algorithm of BN.  The next subsection will be devoted to the explanation of the imputation procedure.

\subsection{Imputing missing values}

\begin{figure}
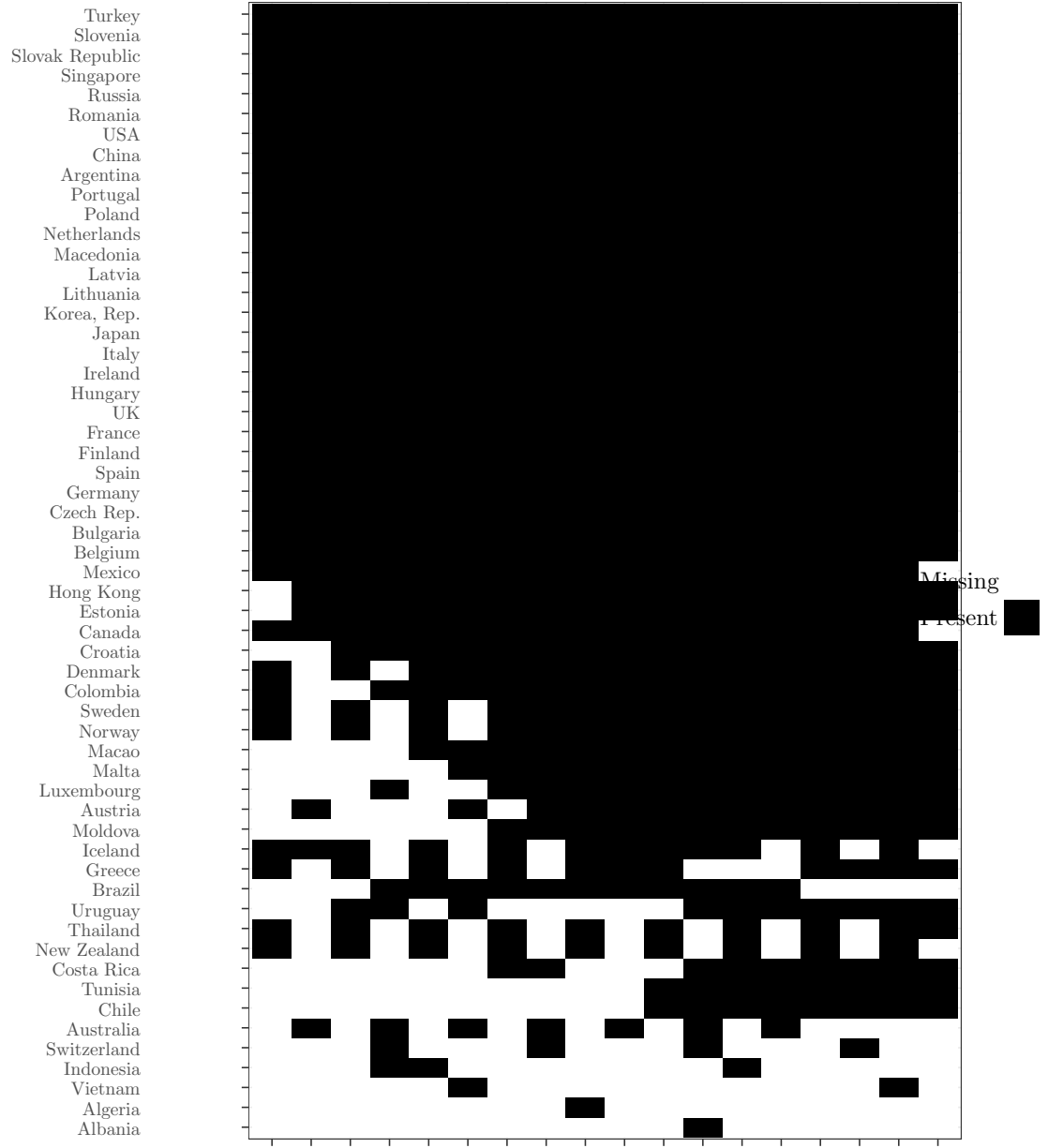



\caption{Representation of the missing values in the data $\mathcal{X}$\label{missingVA}}
\end{figure}

The investment data $\mathcal{X}$ described above and collected from the WDI website  contains some  missing values due to non-collected data. In the data $\mathcal{X}$ we have about 205 missing observations  among 1057 ones. The years 1997 and 1998 have the highest numbers of missing values. It is about 19 and 20 missing values among the 57 countries.   Algeria, Albania and Vietnam are the countries who have  provided the highest numbers of missing values. You can learn more about the missing values in the data from the chart of  Figure \ref{missingVA}.

Many techniques in statistics exist when we want to impute missing values in multivariate data but there exists one method based on the estimation of Bayesian Networks and it is implemented in the command \texttt{impute} in the package \texttt{bnlearn}. This technique works like a prediction of new observations by using the observations of the parents of the missing observations, the estimated BN and its fitted parameters. This method seems suitable in our case study but at this stage we have not yet estimated  the BN and we can  not do so since  the estimation of the BN  precisely requires the dataset without any missing values. This problem looks  like the egg-and-chicken problem.

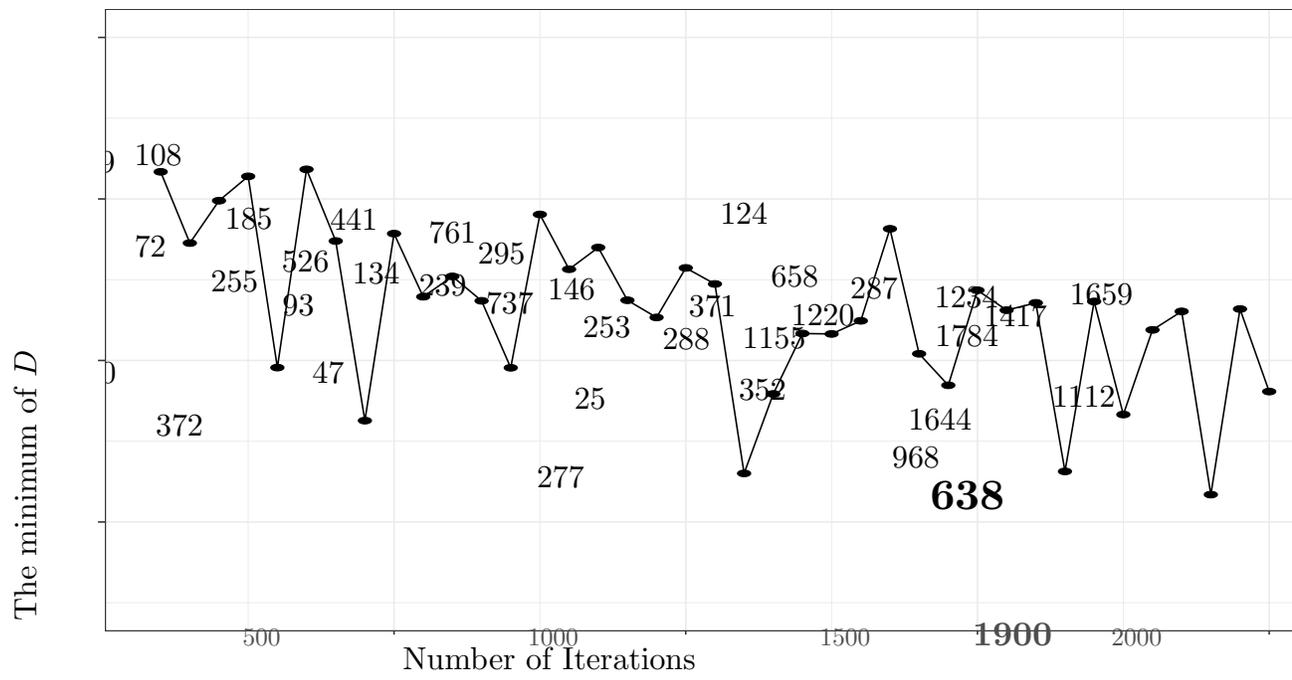
\begin{figure}
\centering
\begin{tikzpicture}[x=1pt,y=.5pt]
\hspace{-2cm}

\definecolor{fillColor}{RGB}{255,255,255}
\path[use as bounding box,fill=fillColor,fill opacity=0.00] (0,0) rectangle (505.89,505.89);
\begin{scope}
\path[clip] (  0.00,  0.00) rectangle (505.89,505.89);
\definecolor{drawColor}{RGB}{255,255,255}
\definecolor{fillColor}{RGB}{255,255,255}

\path[draw=drawColor,line width= 0.6pt,line join=round,line cap=round,fill=fillColor] (  0.00,  0.00) rectangle (505.89,505.89);
\end{scope}
\begin{scope}
\path[clip] ( 39.17, 29.59) rectangle (500.39,500.39);
\definecolor{fillColor}{RGB}{255,255,255}

\path[fill=fillColor] ( 39.17, 29.59) rectangle (500.39,500.39);
\definecolor{drawColor}{gray}{0.92}

\path[draw=drawColor,line width= 0.3pt,line join=round] ( 39.17, 50.99) --
	(500.39, 50.99);

\path[draw=drawColor,line width= 0.3pt,line join=round] ( 39.17,173.27) --
	(500.39,173.27);

\path[draw=drawColor,line width= 0.3pt,line join=round] ( 39.17,295.56) --
	(500.39,295.56);

\path[draw=drawColor,line width= 0.3pt,line join=round] ( 39.17,417.85) --
	(500.39,417.85);

\path[draw=drawColor,line width= 0.3pt,line join=round] ( 93.23, 29.59) --
	( 93.23,500.39);

\path[draw=drawColor,line width= 0.3pt,line join=round] (203.57, 29.59) --
	(203.57,500.39);

\path[draw=drawColor,line width= 0.3pt,line join=round] (313.91, 29.59) --
	(313.91,500.39);

\path[draw=drawColor,line width= 0.3pt,line join=round] (424.26, 29.59) --
	(424.26,500.39);

\path[draw=drawColor,line width= 0.6pt,line join=round] ( 39.17,112.13) --
	(500.39,112.13);

\path[draw=drawColor,line width= 0.6pt,line join=round] ( 39.17,234.42) --
	(500.39,234.42);

\path[draw=drawColor,line width= 0.6pt,line join=round] ( 39.17,356.70) --
	(500.39,356.70);

\path[draw=drawColor,line width= 0.6pt,line join=round] ( 39.17,478.99) --
	(500.39,478.99);

\path[draw=drawColor,line width= 0.6pt,line join=round] (148.40, 29.59) --
	(148.40,500.39);

\path[draw=drawColor,line width= 0.6pt,line join=round] (258.74, 29.59) --
	(258.74,500.39);

\path[draw=drawColor,line width= 0.6pt,line join=round] (369.08, 29.59) --
	(369.08,500.39);

\path[draw=drawColor,line width= 0.6pt,line join=round] (479.43, 29.59) --
	(479.43,500.39);
\definecolor{drawColor}{RGB}{0,0,0}

\path[draw=drawColor,line width= 0.6pt,line join=round] ( 60.13,377.22) --
	( 71.17,323.31) --
	( 82.20,355.38) --
	( 93.23,373.80) --
	(104.27,229.02) --
	(115.30,379.10) --
	(126.34,324.86) --
	(137.37,188.87) --
	(148.40,330.44) --
	(159.44,282.66) --
	(170.47,298.05) --
	(181.51,279.58) --
	(192.54,228.88) --
	(203.57,344.95) --
	(214.61,303.47) --
	(225.64,319.93) --
	(236.68,280.00) --
	(247.71,266.98) --
	(258.74,304.49) --
	(269.78,292.31) --
	(280.81,148.91) --
	(291.85,209.08) --
	(302.88,254.82) --
	(313.91,254.50) --
	(324.95,264.44) --
	(335.98,334.08) --
	(347.02,239.48) --
	(358.05,215.63) --
	(369.08,287.57) --
	(380.12,272.49) --
	(391.15,278.03) --
	(402.19,150.42) --
	(413.22,279.07) --
	(424.26,193.48) --
	(435.29,257.56) --
	(446.32,271.56) --
	(457.36,132.85) --
	(468.39,273.35) --
	(479.43,210.87);
\definecolor{fillColor}{RGB}{0,0,0}

\path[draw=drawColor,line width= 0.4pt,line join=round,line cap=round,fill=fillColor] ( 60.13,377.22) circle (  2.50);

\path[draw=drawColor,line width= 0.4pt,line join=round,line cap=round,fill=fillColor] ( 71.17,323.31) circle (  2.50);

\path[draw=drawColor,line width= 0.4pt,line join=round,line cap=round,fill=fillColor] ( 82.20,355.38) circle (  2.50);

\path[draw=drawColor,line width= 0.4pt,line join=round,line cap=round,fill=fillColor] ( 93.23,373.80) circle (  2.50);

\path[draw=drawColor,line width= 0.4pt,line join=round,line cap=round,fill=fillColor] (104.27,229.02) circle (  2.50);

\path[draw=drawColor,line width= 0.4pt,line join=round,line cap=round,fill=fillColor] (115.30,379.10) circle (  2.50);

\path[draw=drawColor,line width= 0.4pt,line join=round,line cap=round,fill=fillColor] (126.34,324.86) circle (  2.50);

\path[draw=drawColor,line width= 0.4pt,line join=round,line cap=round,fill=fillColor] (137.37,188.87) circle (  2.50);

\path[draw=drawColor,line width= 0.4pt,line join=round,line cap=round,fill=fillColor] (148.40,330.44) circle (  2.50);

\path[draw=drawColor,line width= 0.4pt,line join=round,line cap=round,fill=fillColor] (159.44,282.66) circle (  2.50);

\path[draw=drawColor,line width= 0.4pt,line join=round,line cap=round,fill=fillColor] (170.47,298.05) circle (  2.50);

\path[draw=drawColor,line width= 0.4pt,line join=round,line cap=round,fill=fillColor] (181.51,279.58) circle (  2.50);

\path[draw=drawColor,line width= 0.4pt,line join=round,line cap=round,fill=fillColor] (192.54,228.88) circle (  2.50);

\path[draw=drawColor,line width= 0.4pt,line join=round,line cap=round,fill=fillColor] (203.57,344.95) circle (  2.50);

\path[draw=drawColor,line width= 0.4pt,line join=round,line cap=round,fill=fillColor] (214.61,303.47) circle (  2.50);

\path[draw=drawColor,line width= 0.4pt,line join=round,line cap=round,fill=fillColor] (225.64,319.93) circle (  2.50);

\path[draw=drawColor,line width= 0.4pt,line join=round,line cap=round,fill=fillColor] (236.68,280.00) circle (  2.50);

\path[draw=drawColor,line width= 0.4pt,line join=round,line cap=round,fill=fillColor] (247.71,266.98) circle (  2.50);

\path[draw=drawColor,line width= 0.4pt,line join=round,line cap=round,fill=fillColor] (258.74,304.49) circle (  2.50);

\path[draw=drawColor,line width= 0.4pt,line join=round,line cap=round,fill=fillColor] (269.78,292.31) circle (  2.50);

\path[draw=drawColor,line width= 0.4pt,line join=round,line cap=round,fill=fillColor] (280.81,148.91) circle (  2.50);

\path[draw=drawColor,line width= 0.4pt,line join=round,line cap=round,fill=fillColor] (291.85,209.08) circle (  2.50);

\path[draw=drawColor,line width= 0.4pt,line join=round,line cap=round,fill=fillColor] (302.88,254.82) circle (  2.50);

\path[draw=drawColor,line width= 0.4pt,line join=round,line cap=round,fill=fillColor] (313.91,254.50) circle (  2.50);

\path[draw=drawColor,line width= 0.4pt,line join=round,line cap=round,fill=fillColor] (324.95,264.44) circle (  2.50);

\path[draw=drawColor,line width= 0.4pt,line join=round,line cap=round,fill=fillColor] (335.98,334.08) circle (  2.50);

\path[draw=drawColor,line width= 0.4pt,line join=round,line cap=round,fill=fillColor] (347.02,239.48) circle (  2.50);

\path[draw=drawColor,line width= 0.4pt,line join=round,line cap=round,fill=fillColor] (358.05,215.63) circle (  2.50);

\path[draw=drawColor,line width= 0.4pt,line join=round,line cap=round,fill=fillColor] (369.08,287.57) circle (  2.50);

\path[draw=drawColor,line width= 0.4pt,line join=round,line cap=round,fill=fillColor] (380.12,272.49) circle (  2.50);

\path[draw=drawColor,line width= 0.4pt,line join=round,line cap=round,fill=fillColor] (391.15,278.03) circle (  2.50);

\path[draw=drawColor,line width= 0.4pt,line join=round,line cap=round,fill=fillColor] (402.19,150.42) circle (  2.50);

\path[draw=drawColor,line width= 0.4pt,line join=round,line cap=round,fill=fillColor] (413.22,279.07) circle (  2.50);

\path[draw=drawColor,line width= 0.4pt,line join=round,line cap=round,fill=fillColor] (424.26,193.48) circle (  2.50);

\path[draw=drawColor,line width= 0.4pt,line join=round,line cap=round,fill=fillColor] (435.29,257.56) circle (  2.50);

\path[draw=drawColor,line width= 0.4pt,line join=round,line cap=round,fill=fillColor] (446.32,271.56) circle (  2.50);

\path[draw=drawColor,line width= 0.4pt,line join=round,line cap=round,fill=fillColor] (457.36,132.85) circle (  2.50);

\path[draw=drawColor,line width= 0.4pt,line join=round,line cap=round,fill=fillColor] (468.39,273.35) circle (  2.50);

\path[draw=drawColor,line width= 0.4pt,line join=round,line cap=round,fill=fillColor] (479.43,210.87) circle (  2.50);

\node[text=drawColor,anchor=base,inner sep=0pt, outer sep=0pt, scale=  1.10] at ( 52.78,365.32) {61};

\node[text=drawColor,anchor=base,inner sep=0pt, outer sep=0pt, scale=  1.10] at ( 63.77,311.96) {111};

\node[text=drawColor,anchor=base,inner sep=0pt, outer sep=0pt, scale=  1.10] at ( 74.80,343.32) {78};

\node[text=drawColor,anchor=base,inner sep=0pt, outer sep=0pt, scale=  1.10] at ( 99.50,376.90) {49};

\node[text=drawColor,anchor=base,inner sep=0pt, outer sep=0pt, scale=  1.10] at ( 96.97,217.24) {180};

\node[text=drawColor,anchor=base,inner sep=0pt, outer sep=0pt, scale=  1.10] at (121.85,382.33) {108};

\node[text=drawColor,anchor=base,inner sep=0pt, outer sep=0pt, scale=  1.10] at (118.77,313.33) {72};

\node[text=drawColor,anchor=base,inner sep=0pt, outer sep=0pt, scale=  1.10] at (129.86,177.41) {372};

\node[text=drawColor,anchor=base,inner sep=0pt, outer sep=0pt, scale=  1.10] at (156.10,334.51) {185};

\node[text=drawColor,anchor=base,inner sep=0pt, outer sep=0pt, scale=  1.10] at (150.68,286.64) {255};

\node[text=drawColor,anchor=base,inner sep=0pt, outer sep=0pt, scale=  1.10] at (177.51,301.42) {526};

\node[text=drawColor,anchor=base,inner sep=0pt, outer sep=0pt, scale=  1.10] at (174.80,268.94) {93};

\node[text=drawColor,anchor=base,inner sep=0pt, outer sep=0pt, scale=  1.10] at (186.09,217.77) {47};

\node[text=drawColor,anchor=base,inner sep=0pt, outer sep=0pt, scale=  1.10] at (195.83,333.25) {441};

\node[text=drawColor,anchor=base,inner sep=0pt, outer sep=0pt, scale=  1.10] at (204.27,292.61) {134};

\node[text=drawColor,anchor=base,inner sep=0pt, outer sep=0pt, scale=  1.10] at (233.10,323.35) {761};

\node[text=drawColor,anchor=base,inner sep=0pt, outer sep=0pt, scale=  1.10] at (229.41,283.36) {239};

\node[text=drawColor,anchor=base,inner sep=0pt, outer sep=0pt, scale=  1.10] at (254.79,270.41) {737};

\node[text=drawColor,anchor=base,inner sep=0pt, outer sep=0pt, scale=  1.10] at (251.67,307.84) {295};

\node[text=drawColor,anchor=base,inner sep=0pt, outer sep=0pt, scale=  1.10] at (278.20,280.61) {146};

\node[text=drawColor,anchor=base,inner sep=0pt, outer sep=0pt, scale=  1.10] at (273.97,138.14) {277};

\node[text=drawColor,anchor=base,inner sep=0pt, outer sep=0pt, scale=  1.10] at (285.13,197.68) {25};

\node[text=drawColor,anchor=base,inner sep=0pt, outer sep=0pt, scale=  1.10] at (291.50,252.44) {253};

\node[text=drawColor,anchor=base,inner sep=0pt, outer sep=0pt, scale=  1.10] at (321.52,243.34) {288};

\node[text=drawColor,anchor=base,inner sep=0pt, outer sep=0pt, scale=  1.10] at (331.52,268.01) {371};

\node[text=drawColor,anchor=base,inner sep=0pt, outer sep=0pt, scale=  1.10] at (343.38,337.85) {124};

\node[text=drawColor,anchor=base,inner sep=0pt, outer sep=0pt, scale=  1.10] at (354.84,243.63) {1155};

\node[text=drawColor,anchor=base,inner sep=0pt, outer sep=0pt, scale=  1.10] at (350.39,204.78) {352};

\node[text=drawColor,anchor=base,inner sep=0pt, outer sep=0pt, scale=  1.10] at (362.62,290.66) {658};

\node[text=drawColor,anchor=base,inner sep=0pt, outer sep=0pt, scale=  1.10] at (373.25,261.21) {1220};

\node[text=drawColor,anchor=base,inner sep=0pt, outer sep=0pt, scale=  1.10] at (392.59,281.55) {287};

\node[text=drawColor,anchor=base,inner sep=0pt, outer sep=0pt, scale=  1.10] at (408.42,153.67) {968};

\node[text=drawColor,anchor=base,inner sep=0pt, outer sep=0pt, scale=  1.10] at (427.46,274.37) {1234};

\node[text=drawColor,anchor=base,inner sep=0pt, outer sep=0pt, scale=  1.10] at (417.57,182.11) {1644};

\node[text=drawColor,anchor=base,inner sep=0pt, outer sep=0pt, scale=  1.10] at (427.75,245.60) {1784};

\node[text=drawColor,anchor=base,inner sep=0pt, outer sep=0pt, scale=  1.10] at (446.12,260.71) {1417};

\node[text=drawColor,anchor=base,inner sep=0pt, outer sep=0pt, scale=  1.50] at (450.63,122.17) {\textbf{638}};

\node[text=drawColor,anchor=base,inner sep=0pt, outer sep=0pt, scale=  1.10] at (478.62,276.82) {1659};

\node[text=drawColor,anchor=base,inner sep=0pt, outer sep=0pt, scale=  1.10] at (471.95,199.43) {1112};
\definecolor{drawColor}{gray}{0.20}

\path[draw=drawColor,line width= 0.6pt,line join=round,line cap=round] ( 39.17, 29.59) rectangle (500.39,500.39);
\end{scope}
\begin{scope}
\path[clip] (  0.00,  0.00) rectangle (505.89,505.89);
\definecolor{drawColor}{gray}{0.30}

\node[text=drawColor,anchor=base east,inner sep=0pt, outer sep=0pt, scale=  0.88] at ( 34.22,109.10) {0.10};

\node[text=drawColor,anchor=base east,inner sep=0pt, outer sep=0pt, scale=  0.88] at ( 34.22,231.39) {0.15};

\node[text=drawColor,anchor=base east,inner sep=0pt, outer sep=0pt, scale=  0.88] at ( 34.22,353.67) {0.20};

\node[text=drawColor,anchor=base east,inner sep=0pt, outer sep=0pt, scale=  0.88] at ( 34.22,475.96) {0.25};
\end{scope}
\begin{scope}
\path[clip] (  0.00,  0.00) rectangle (505.89,505.89);
\definecolor{drawColor}{gray}{0.20}

\path[draw=drawColor,line width= 0.6pt,line join=round] ( 36.42,112.13) --
	( 39.17,112.13);

\path[draw=drawColor,line width= 0.6pt,line join=round] ( 36.42,234.42) --
	( 39.17,234.42);

\path[draw=drawColor,line width= 0.6pt,line join=round] ( 36.42,356.70) --
	( 39.17,356.70);

\path[draw=drawColor,line width= 0.6pt,line join=round] ( 36.42,478.99) --
	( 39.17,478.99);
\end{scope}
\begin{scope}
\path[clip] (  0.00,  0.00) rectangle (505.89,505.89);
\definecolor{drawColor}{gray}{0.20}

\path[draw=drawColor,line width= 0.6pt,line join=round] (148.40, 26.84) --
	(148.40, 29.59);

\path[draw=drawColor,line width= 0.6pt,line join=round] (258.74, 26.84) --
	(258.74, 29.59);

\path[draw=drawColor,line width= 0.6pt,line join=round] (369.08, 26.84) --
	(369.08, 29.59);

\path[draw=drawColor,line width= 0.6pt,line join=round] (479.43, 26.84) --
	(479.43, 29.59);
\end{scope}
\begin{scope}
\path[clip] (  0.00,  0.00) rectangle (505.89,505.89);
\definecolor{drawColor}{gray}{0.30}

\node[text=drawColor,anchor=base,inner sep=0pt, outer sep=0pt, scale=  0.88] at (148.40, 18.58) {500};

\node[text=drawColor,anchor=base,inner sep=0pt, outer sep=0pt, scale=  0.88] at (258.74, 18.58) {1000};

\node[text=drawColor,anchor=base,inner sep=0pt, outer sep=0pt, scale=  0.88] at (369.08, 18.58) {1500};

\node[text=drawColor,anchor=base,inner sep=0pt, outer sep=0pt, scale=  1.20] at (450.63,18.58) {\textbf{1900}};

\node[text=drawColor,anchor=base,inner sep=0pt, outer sep=0pt, scale=  0.88] at (479.43, 18.58) {2000};
\end{scope}
\begin{scope}
\path[clip] (  0.00,  0.00) rectangle (505.89,505.89);
\definecolor{drawColor}{RGB}{0,0,0}

\node[text=drawColor,anchor=base,inner sep=0pt, outer sep=0pt, scale=  1.10] at (269.78,  0.50) {Number of Iterations};
\end{scope}
\begin{scope}
\path[clip] (  0.00,  0.00) rectangle (505.89,505.89);
\definecolor{drawColor}{RGB}{0,0,0}

\node[text=drawColor,rotate= 90.00,anchor=base,inner sep=0pt, outer sep=0pt, scale=  1.10] at ( 13.08,264.99) {The minimum of $D$};
\end{scope}
\end{tikzpicture}

\caption{Result of the BNII.}  \label{BII}
\end{figure}

To solve the problem we set up a new procedure based on an iterative method  giving an  imputation in our opinion better than the existing imputation methods and we call the obtained  algorithm   Bayesian Network Iterative Imputation algorithm (BNII).   We   mimic the  famous cross-validation method used in classification and regression  methods to provide an estimation of the predicted error. At each step of the iteration, we drop randomly real observations, make them missing, impute them using BN and measure the gap between the new imputed values and  their real observed values as defined in (\ref{gapE}). We  stop the procedure when we estimate that we  reach a reasonable low value of this gap. In practice we have run this procedure from 100 to 2000 times by considering steps of 50 iterations. In each iteration we record the minimum of the gap $D$ defined in (\ref{gapE}) and its rank in the iteration. We  summarize this result in Figure \ref{BII}, where we put on the $x-$axis the number of iterations and $D$ on the $y-$axis and where the  annotated integer   is the rank of the observed minimum at the corresponding iteration.


We can notice that $D$ increases slowly when the number of iterations increases and its minimum can be reached before the end of the fixed number of iterations. We   observe that this minimum is reached when  the procedure is run 1900 times and the minimum is observed at the 638-th iteration.      

  An initial step  consists in an imputation using the $K−$Nearest
Neighbors algorithm (KNN). The idea behind using KNN for missing
values is that the \textit{investment} in R\&D can be approximated by the other \textit{investment} observations that are closest to it. In this procedure we  replace the missing value by the median of the 10 closest neighbors. We obtain then an initial imputed data $\mathcal{X}_0^c$.

Then we  start our iterative procedure. First  the BN is estimated from the completed data from the previous step. Then we remove randomly $n=50$ observations.  Using the fitted model we estimate the missing values  to complete  the data and measure $D$. We continue this iteration for a fixed number of times. The whole procedure is  described in Algorithm \ref{bnii} 
where 

\begin{itemize}

\item $\mathcal{S}_{T}$ is the whole set of  indexes  of the   observations in the data $\mathcal{X}$, i.e. $\mathcal{S}_{T}=\{1,\ldots,n\}\times \{1,\ldots,p\} $, 

\item $n$ and $p$ are respectively the number of rows and columns in $\mathcal{X}$, 

\item $\mathcal{S}_{NA}$ is the set of indexes  of the  missing observations in the data $\mathcal{X}$, 
$$\mathcal{S}_{NA}\subset \mathcal{S}_{T}.$$

\end{itemize}

\begin{algorithm}

\caption{Bayesian Network Iterative Imputation Algorithm}\label{bnii}
\begin{algorithmic}[1]

\State let $i=0$. 
\State Start with $\mathcal{X}_i$ a complete version of $\mathcal{X}$ computed using KNN algorithm. 
\State let $N$ be  the maximum number of iterations for the coming loop. 
\State Let fix $d_i>0$  an initial large positive number.    
\While{$i\leq N$}
\State Sample  from $\mathcal{S}_T\setminus \mathcal{S}_{NA}$ $50$ indexes   complete observations from $\mathcal{X}$ and transform them to  missing values. Let's denote by $\mathcal{S}_r$ this set of indexes, i.e; $ \mathcal{S}_r\subset \mathcal{S}_T\setminus \mathcal{S}_{NA}$
\State Denote by $\mathcal{X}'$ the new version of $\mathcal{X}$ with missing values at $\mathcal{S}_r\cup \mathcal{S}_{NA}$.
\State Estimate and fit the BN  using $HC$ algorithm from $\mathcal{X}_i$. Let's denote by $\widehat{G}$ the estimated BN model.
\State Impute $\mathcal{X}'$ using $\widehat{G}$ and obtain a new complete version of $\mathcal{X}$. Let's  denote it by $\mathcal{X}_c$.
\State Compute 
\begin{equation}
\label{gapE}
D=\sum_{s\in \mathcal{S}_r} (x(s)-x_c(s))^2
\end{equation} 
where $x(s)$ and $x_c(s)$ are respectively is the generic coefficient  of $\mathcal{X}$ and $\mathcal{X}_c$.

\If {$d_i \geq  D$} 
\State $i\leftarrow i+1$, $d_i\leftarrow D$ and $\mathcal{X}_i=\mathcal{X}_c$. 
\Else 
\State $i\leftarrow N+1$
\EndIf
 \EndWhile
\State \textbf{return} $\mathcal{X}_i$ a complete version of the data
\end{algorithmic}
\end{algorithm}


\section{Results}
\begin{figure}
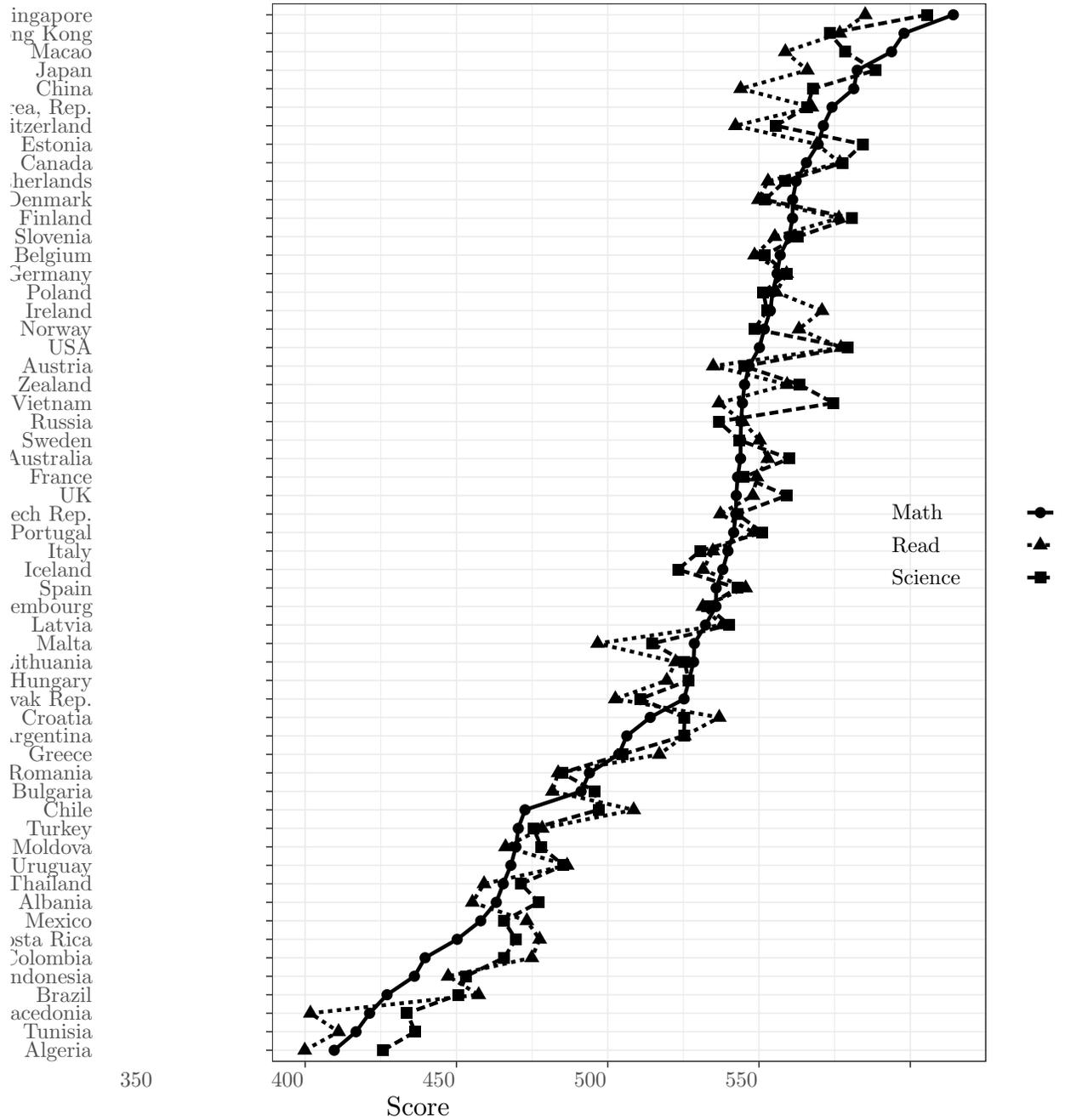

\centering


\caption{PISA scores\label{p-scores}}
\end{figure}

\begin{figure}
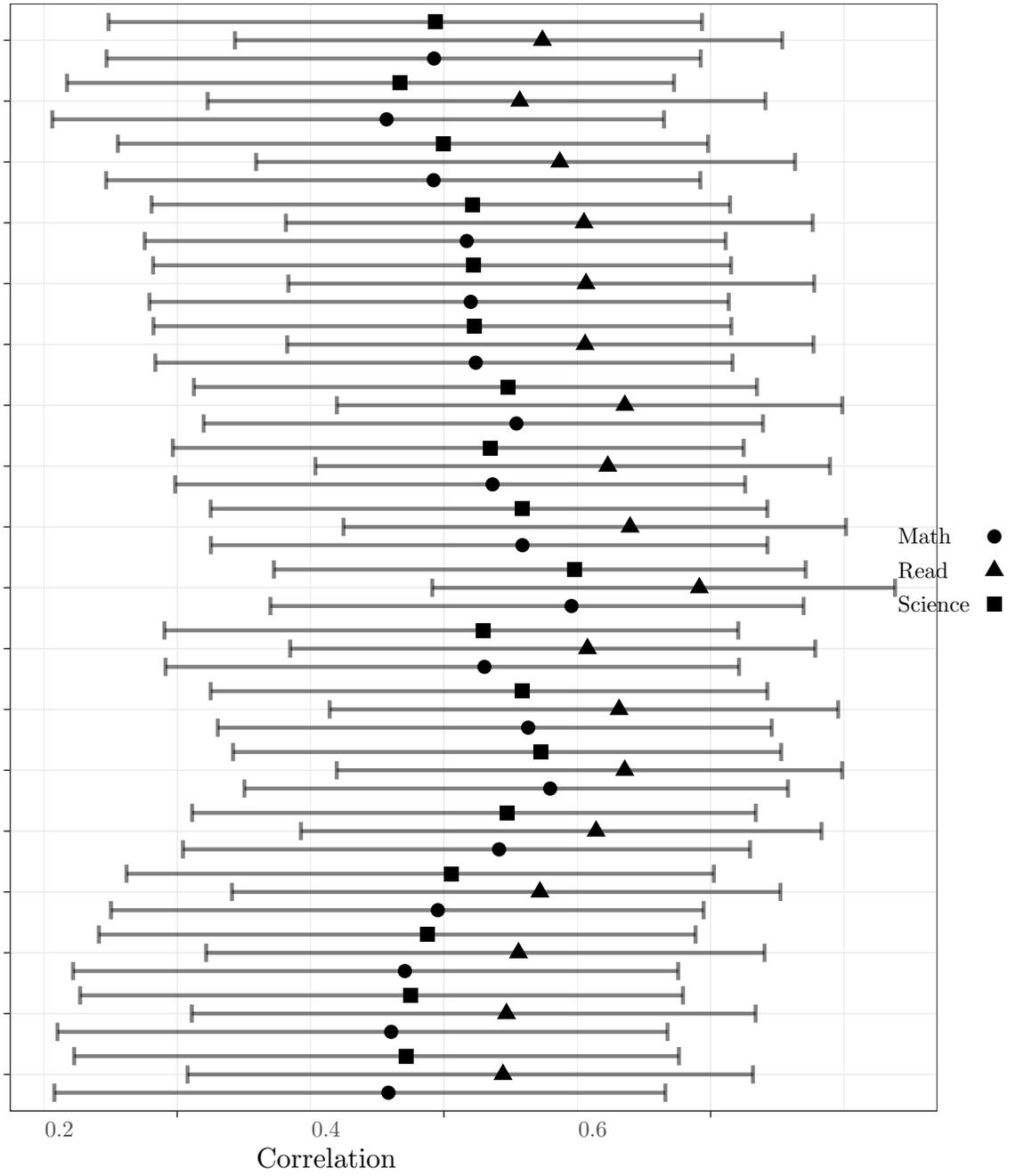

\centering
%

\caption{Correlation and their confidence intervals of PISA scores with \textit{Investment} variables from 1997 to 2014\label{cor-scores}}
\end{figure}

\vskip 5 truemm
We  start this section with two charts (Figures \ref{p-scores} and \ref{cor-scores}). In Figure \ref{p-scores} we represent the countries in our data with respect to their three PISA Scores: Math, Reading and Science. We sort them from the less performing: Algeria and Tunisia to the most performing countries Singapore, Hong Kong, Macao and Japan.  We  notice from this graph the strong correlation between these three scores and we can even conclude that they are redundant. This is why  we have decided to consider   only the ``Reading'' score to understand the impact of the investment in R\&D on the performance of Education.

Figure \ref{cor-scores} is a representation of the correlation of the PISA scores with each of the investment variables in R\&D from 1997 to 2014. We also represent with segments the 95\% confidence interval (CI) of each correlation coefficient. First we notice that none of these CIs contains zero. This means that the relationship between investment in R\&D and performance in Education is always significant. Hence whenever you invest in R\&D there is always a significant impact on the performance of Education. One important question is when this investment has the highest impact on Education, i.e. is the effect  stronger in the short or the long run? This is the reason of our use  of a sophisticated model such as the Bayesian Networks: to make sure that we are computing properly  the effects of the investments in R\&D of all the preceding years on Education and be able to compare them.  The years from 2002 to 2005 seem to have slightly higher correlation than all the other years. 
Does the estimation of Bayesian Network  confirm this preliminary  observation?

\begin{figure}
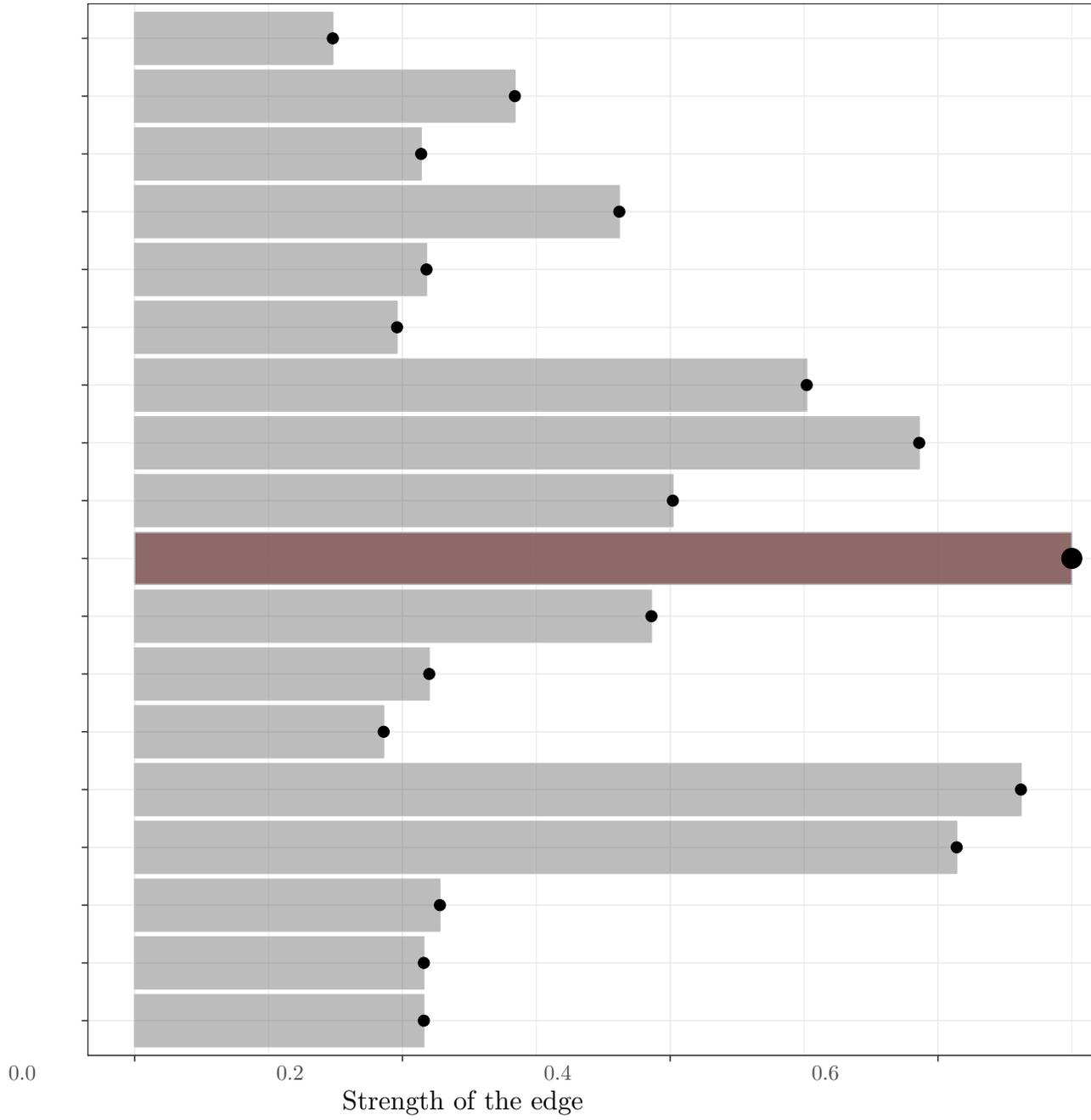

\centering


\caption{Strength of the link R\&D and Reading Score\label{StrengthRead}}
\end{figure}

We  run 500 times the bootstrap procedure to estimate the Bayesian Network. We represent in Figure \ref{StrengthRead} the strength of each arrow pointing from each investment year to the Education variable ``Reading". We can easily notice that the investment  the most likely linked to the ``Reading'' variable is the one made in 2005. For example the arrow linking the variables representing the investment in R\&D in 2005  and ``Reading''  is present in 70\% of the estimated BN. The arrow pointing from  R\&D in 2014 to  ``Reading'' is only present in 14\% of the estimated BN amid the 500 bootstrap replicates. This fact shows obviously that the highest impact on the performance of  Education is recorded from the investment in R\&D 10 years earlier. 

\begin{figure}
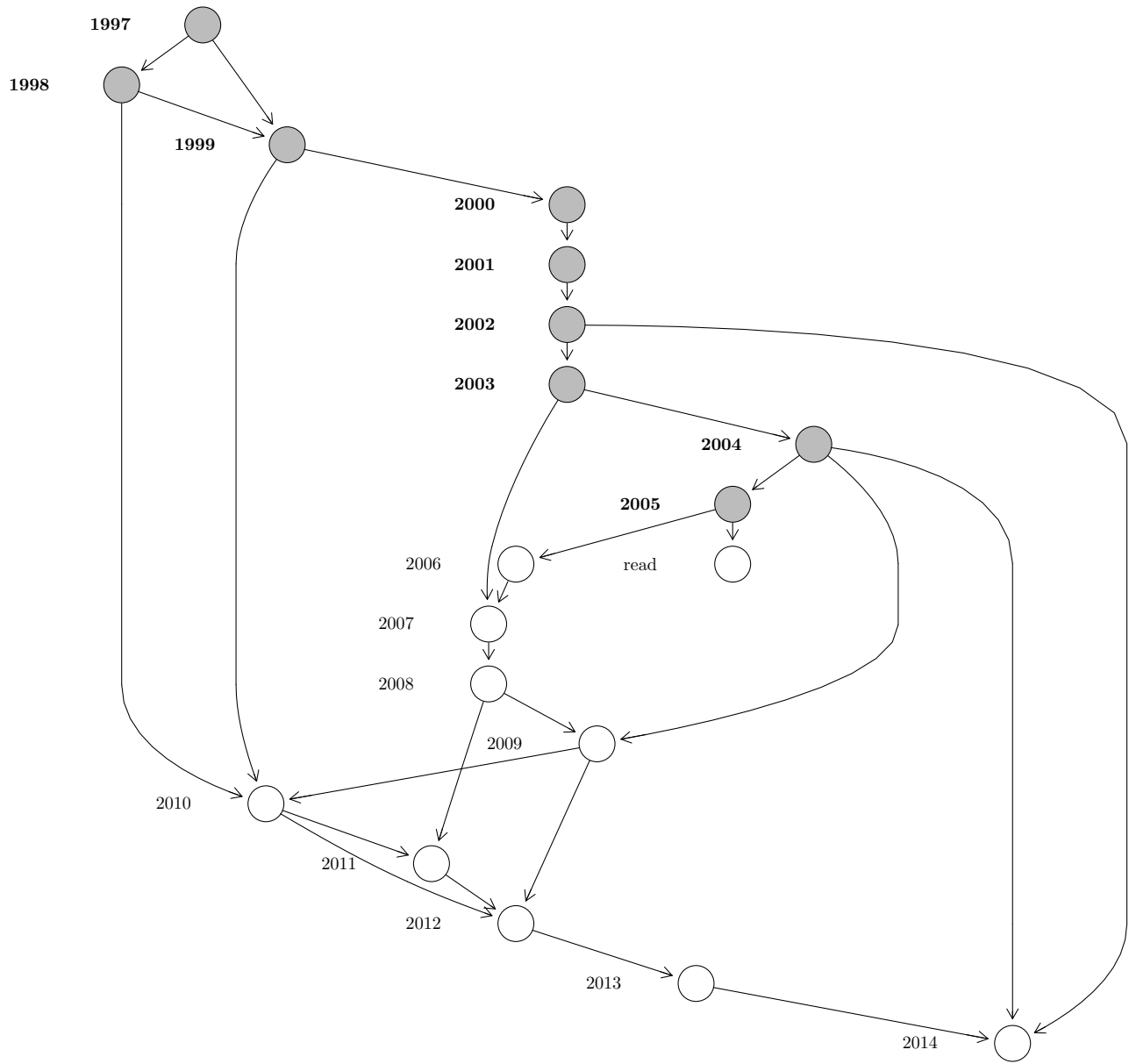

\centering
%

\caption{Estimated DBN for Reading Score by highlighting the path to the Math-variable\label{DBNRead}}
\end{figure}

We have  chosen a threshold equal to 60\% when computing the  average of the models among the 500 estimated during the bootstrap procedure. The estimated BN is then represented in Figure \ref{DBNRead}. From this BN we can notice that  the investment in 1997 affects the one in 1998 and 1999 and this impact will be successively continued until 2005 which is the last year that impacts highly the performance ``Reading".

We have also estimated the parameters in the models by successively performing a series of linear regression models according to the path from the investment variable in 1997 to the PISA scores. This path is colored in gray in the representation of the estimated final BN in Figure \ref{DBNRead}. These estimations are displayed in Table \ref{regressions}. In this table we show the estimation of the parameters of each model determined by each node in the path between the variable representing the investment in 1997 to the ``Reading'' PISA score. If $v$ is  a node in that path, the estimated regression shown in Table \ref{regressions} is the one defined by   $\theta(v)$  as  a dependent variable and $\Theta(u)$ as the vector of  independent variables, $u$ is a node parent of $v$, i.e; $(\Theta(u),\, u\in pa(v))$. Note that in Table \ref{regressions}, the dependent variables are on the columns  and the independent are in the rows. Empty cells in Table \ref{regressions} show that the corresponding variable  is not a parent of the dependent variable in the estimated BN.  Hence by performing this series of regression models we can deduce the conditional probability distribution of $\theta(v)\mid  \Theta(u),\, u\in pa(v)$ for every $v$ in the path between the variable investment in R\&D in 1997 to the variable ``Reading". Thus by using the Markov property introduced in (\ref{facMark}) we can  deduce the estimation of the parameters of the probability distribution of  the random vector composed by the variables ``Reading" and those on the path between it and the variable R\&D in 1997.

\subsection{Interpretation}

We are going to explain why there is a causal relationship between expenditure in R\&D and education in small classes and to understand why such a long delay is necessary. 

R\&D is intended to better understand the world and make discoveries meant to improve
human life some way or the other. This is how most people perceive R\&D.
R\&D is also intended to train the trainers, those who will be university teachers (through the supervision of Phd theses) but also the teachers who will educate the children in the primary and secondary schools, the future parents and all types of educators who are likely to take care of the children for diverse activities (sports, excursions, cultural and recreational activities...). One dollar spent in R\&D has repercussions in the short term on the output of research (publications, patents...) and thus on the level of the university teachers. But the latter has an effect on the quality of training of the students among whom the future parents and the educators of children. 
This is in line with the numerous papers who have proved that mother's and father's education positively influences child quality in several respects (health, cognitive development, {\bf education}, occupation status, future earnings): \citeauthor{l74} (\citeyear{l74}, \citeyear{l75}),  \citeauthor{eg79} (\citeyear{eg79}), \citeauthor{bg70} (\citeyear{bg70}), \citeauthor{hs74} (\citeyear{hs74,hs80}), \citeauthor{wb82} (\citeyear{wb82}), among numerous others.

This explanation is confirmed by the lag we measured between the spending in R\&D and the actual improvement of the children's performance in PISA tests. This lag is around 10 years, which seems to be a reasonable delay to obtain the effect in question. Indeed the investment in R\&D has first to affect positively the performance of university teachers, before reaching the education of children. 


This is not to say that good researchers are necessarily good university teachers or that university teachers cannot be good if they do not do enough research. First it is a {\it statistical} causal relationship which may fail to hold in some cases. Second, it is the environment of teaching as a whole which is important in universities. When there is a dynamic research in a university, even the teachers who, strictly speaking, do not do research (not writing articles  or filing patents for instance)   benefit from the ``research environment" where they attend seminars, discuss with colleagues who are researchers, meet invited professors thanks to the activity of researchers, are supervised and/or surrounded by active researchers... The training plans  evolve better under the influence of university teachers ``up-to-date" with the recent discoveries and theories. The research creates an environment favorable to questioning, competition and progress, which is beneficial to teaching practices and all university teachers even those who are not active researchers. This statement is confirmed by our paper. The time necessary to obtain this effect corresponds to the time necessary for the research performance itself to be positively affected by the investment in R\&D, then for this performance to affect positively the quality of teaching and finally to improve the education performance through a better quality of graduates. Research contributes this way highly to boost the universities and the whole  educative system.

\subsection{Other applications}

We may use the obtained estimation to serve a variety of purposes. We first estimate the participation of R\&D in the performance of education. Second we evaluate the efficiency in each country of R\&D in education performance.

\hspace{-7cm}
\begin{table}[!htbp] \centering 

  \caption{Regression models from the best fitted BN} 
  \label{regressions} 
  \begin{adjustbox}{width=1\textwidth}
\small
\begin{tabular}{@{\extracolsep{5pt}}lcccccccc} 
\\[-1.8ex]\hline 
\hline \\[-1.8ex] 
 & \multicolumn{8}{c}{\textit{Dependent variable:}} \\ 
\cline{2-9} 
\\[-1.8ex] & `1998` & `1999` & `2000` & `2001` & `2003` & `2004` & `2005` & Read \\ 
\\[-1.8ex] & (1) & (2) & (3) & (4) & (5) & (6) & (7) & (8)\\ 
\hline \\
\end{tabular} 
\end{adjustbox}
\end{table} 

\subsection{Participation of R\&D in the performance of Education}

 The participation of R\&D in the performance of Education is computed using the estimation of the intercept of regression model where $Y$ is the variable ``Reading'' as a dependent variable  and where $\theta(2005)$  is  the variable investment in 2005 as an independent variable.  The regression model can be written as 
\begin{equation}
\label{regMod}
Y =\alpha + \beta \times  \theta(2005)+\epsilon,
\end{equation}

where  $\epsilon$ is the random error  supposed to follow a central Gaussian distribution with unknown variance $\sigma^2$ and where $\alpha$ and $\beta$ are respectively the intercept and the slope of the regression. According to the result displayed in Table \ref{regressions}, the estimation of $\alpha$ is equal to $\widehat{\alpha}=192.574$ and the estimation of $\beta$ is equal to $\widehat{\beta}=25.239$. 

The first coefficient $\widehat{\alpha}$ can be interpreted as the expected Education score when no investment is done in R\&D. The second coefficient $\widehat{\beta}=25.239$ can be interpreted as the gain in Education score when it is decided to invest 1\% more in  R\&D.  We can conclude that an Education score can gain an average of 25 points when   1\% more is invested in 2005.  The coefficient $\widehat{\alpha}$ can also be interpreted as the average  score in ``Reading" that can not be explained by the investment in R\&D. We  then compute the percentage of performance in Education that can be explained by the investment in R\&D (and call it simply Contribution of R\&D)  as follows:

\begin{equation}
\label{contribRD}
\mbox{Contribution of R\&D}(w)=\displaystyle\frac{Y(w)-\widehat{\alpha}}{Y(w)},
\end{equation}
where, for every country $w$, $Y(w)$ is its ``Reading'' score.  

\begin{figure}
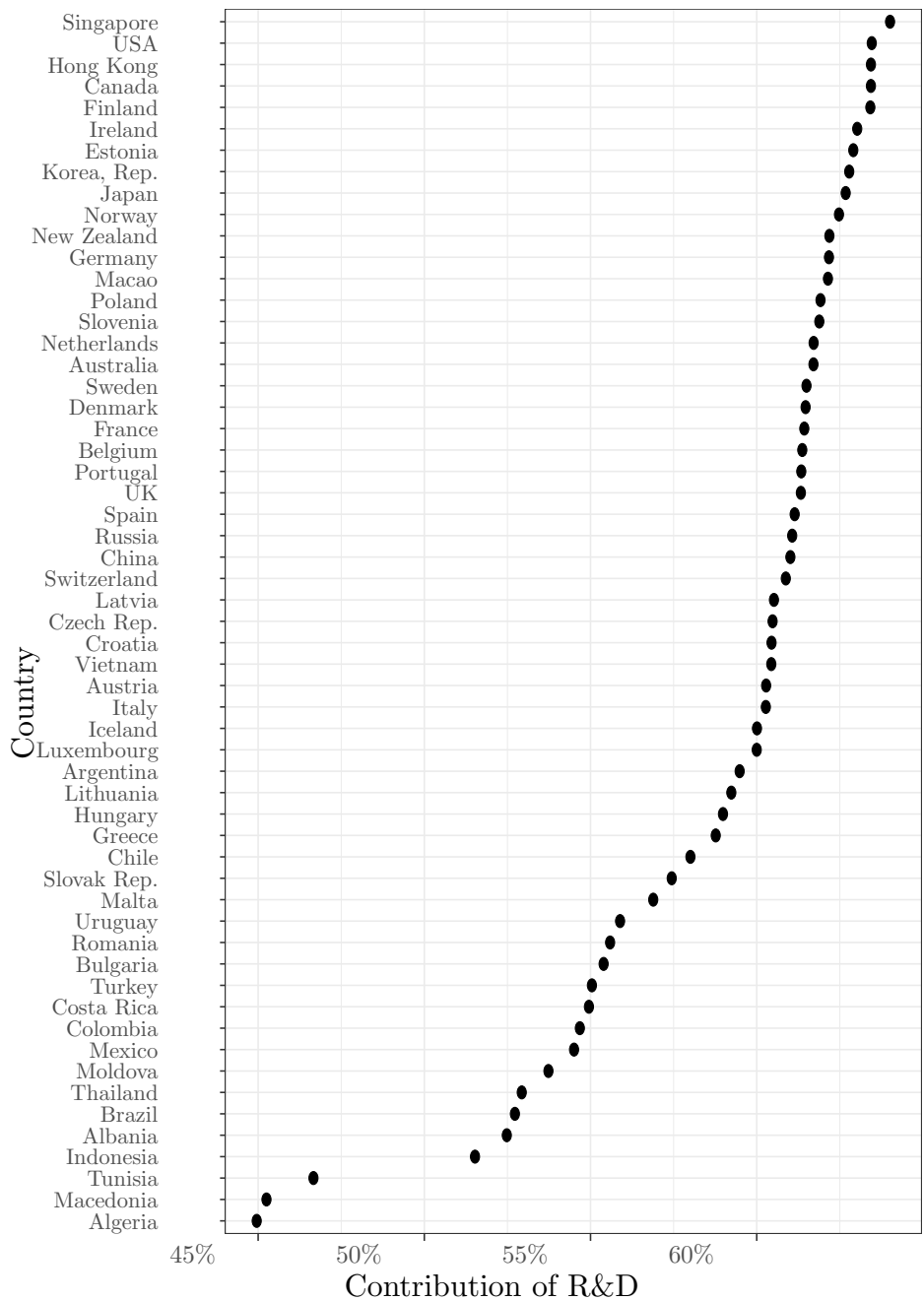

\centering


\caption{Contribution of R\&D in the performance of Education as defined in (\ref{contribRD}) by country\label{partRD}}
\end{figure}

The Contribution of R\&D is computed for every country and depicted in Figure \ref{partRD}. Countries are ranked increasingly with the obtained Contribution of R\&D. The contribution of R\&D in the performance of Education is the weakest in Algeria, Macedonia and Tunisia which also are the last countries in terms of PISA scores. This index ranges from 45\% to 68 \%, which is high. But the difference between different countries may be also high.

\subsection{Efficiency of R\&D in the performance of Education}

\begin{figure}
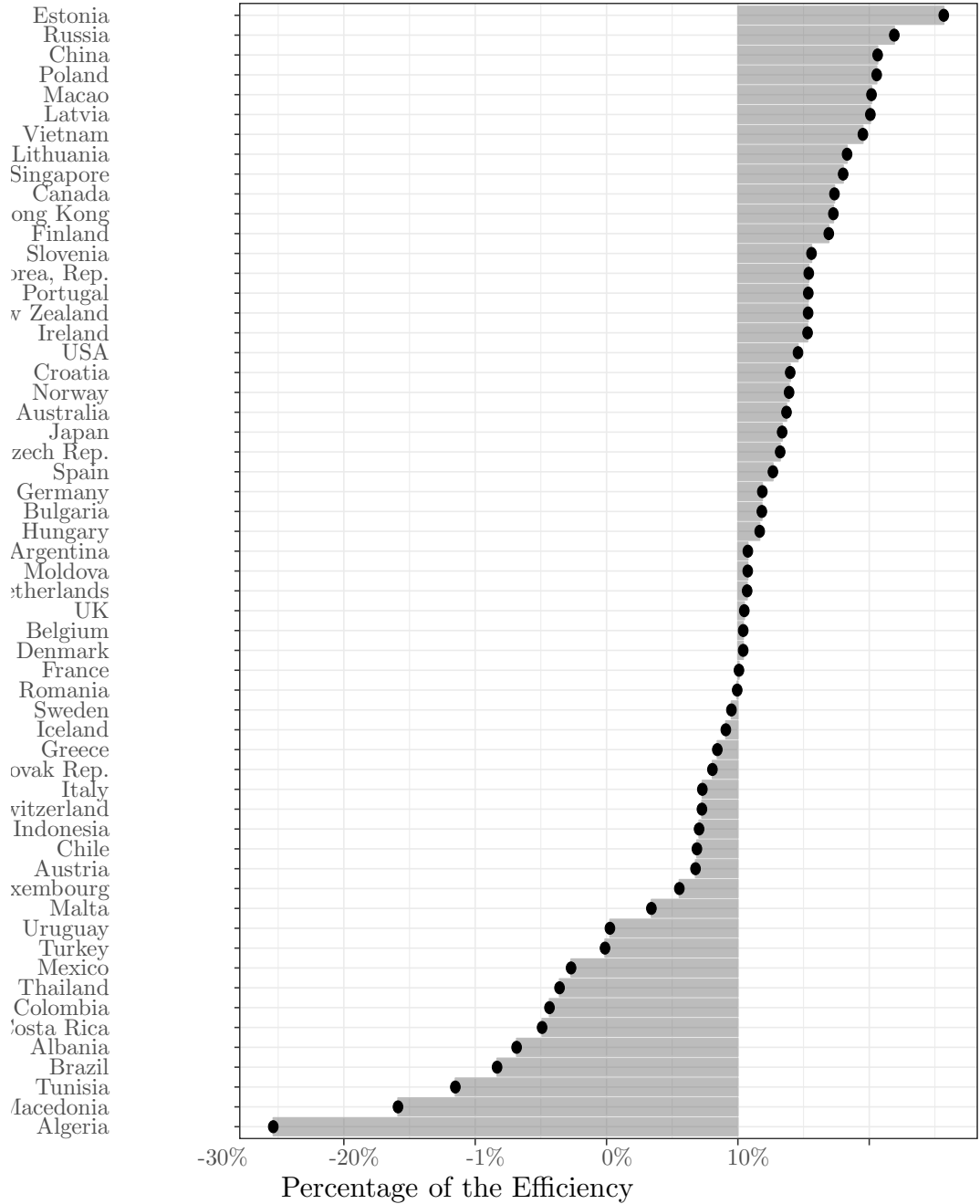

\centering
%

\caption{Efficiency Index defined in (\ref{effiRD}) by country\label{EffRD}}
\end{figure}

A second way to use the BN model is to evaluate the efficiency of R\&D in the performance of Education. This is done by estimating the expected PISA score for each country and comparing it to the observed one.   We define for each country $w$, the Efficiency of R\&D in Education performance as the index $\mbox{EffiRD}(w)$ given by the following: 
\begin{equation}
\label{effiRD}
\mbox{Efficiency of R\&D}(w)=\displaystyle\frac{Y(w)-\widehat{Y}(w)}{Y(w)},
\end{equation}
where $\widehat{Y}(w)$ is the expected score computed using the adjusted BN model and $Y(w)$ is the observed  score. This index, given by the difference between the observed score $Y(w)$ and the expected Education $\widehat{Y}(w)$ divided by the observed score $Y(w)$,  can be interpreted as the Efficiency of the investment of R\&D in the Education performance, as it  shows whether the considered country is more or less efficient than expected and measures this efficiency or lack of efficiency in percentage of the observed performance.  When this index is positive it means that we do better than what is expected and when it is negative we do worse.  In Figure \ref{EffRD} we show a representation of this index. As expected the countries having the worst Education scores have the least Efficient R\&D in Education. We note substantial differences across countries, efficiency of R\&D ranging from -35\% to +15\%. For example,  the Education performance of Tunisia is less by 21\%   than the expected one. But Estonia does 15\% better than what it is expected. Among the 57 considered countries, 23 have negative values of  Efficiency of R\&D in Education performance, suggesting that they can do more with the same invested amounts, or equivalently, that their educative system is somehow deprived from ``free" additional performance.

Considering  together the two new indexes (Contribution of R\&D in the explanation of education performance  and Efficiency of R\&D in education performance) allow to refine the analysis and the conclusions. Indeed the contribution of R\&D in education ranges from 45\% to 64 \%, which is very high. But the differences between countries is also high. These differences may be explained by the differences in efforts and differences in efficiency of R\&D. R\&D contributes more or less in the performance of education because countries invest more or less in R\&D and/or because R\&D is more or less efficient in the performance of education. This is to say that investing more in R\&D may be insufficient, one has also to be efficient to perform better with less. Even if it is not explicit in our modeling, the organization of R\&D and the mechanism of transmission of its effects to university teaching and the quality of graduation appear to be important to obtain the desired improvement in education performance.
\section{Conclusion}
Considering the World Development Indicators and the national PISA scores obtained in 2015 and using Learning Bayesian Networks, we prove that R\&D expenditures affect positively education performance, explaining more than 45\% of the education performance.

Research contributes highly to boost the universities, we already knew that. Our paper provides evidence that it also boosts the whole  educative system. Research is not a luxury devoted to developed countries, and may even be more vital for developing countries where the educative systems may be more fragile. Nor is it  a luxury limited to fat years in developed countries. And claiming every now and then, especially in lean years, that researchers do not make enough discoveries, is not an acceptable reason to justify to cut research budgets.

But this conclusion does not absolve countries and researchers from  the obligation of  trying to make  R\&D  more efficient. Indeed, when measuring the efficiency of R\&D in the performance of R\&D, we also noted substantial differences between countries, suggesting that  more than 23 among the 57 countries can do better with the same invested amounts. A better organization of research and more efforts for the transmission of the benefits of R\&D to university teaching and graduation are as necessary as  investments in R\&D, in order to secure the desired improvement of education performance.





\section*{Bibliography}

\end{document}